%% file: final.tex
\documentclass[prl,twocolumn,showpacs,floatfix,superscriptaddress,nofootinbib,longbibliography]{revtex4-1}

\usepackage{amssymb,amsmath,amstext}                
\usepackage{graphicx}                                               
\usepackage{epstopdf}                                               
\usepackage{color}       
\usepackage{bm}
\usepackage{appendix}                                              
\usepackage[utf8]{inputenc}
\usepackage{bbold}
\usepackage{bbm}
\usepackage{ulem}
\usepackage{latexsym}
\usepackage[colorlinks=true,citecolor=blue,linkcolor=magenta]{hyperref}

\begin{document}
\title{
Universal shift of the fidelity susceptibility peak away from 
the critical point of \\ the Berezinskii-Kosterlitz-Thouless quantum phase transition
}

\author{Lukasz Cincio}
\affiliation{Theory Division, Los Alamos National Laboratory, Los Alamos, NM 87545, USA}
\author{Marek M. Rams}
\affiliation{Institute of Physics, Jagiellonian University,  \L{}ojasiewicza 11, 30-348 Krak\'ow, Poland }
\author{Jacek Dziarmaga}
\affiliation{Institute of Physics, Jagiellonian University,  \L{}ojasiewicza 11, 30-348 Krak\'ow, Poland }
\author{Wojciech H. Zurek}
\affiliation{Theory Division, Los Alamos National Laboratory, Los Alamos, NM 87545, USA}

\begin{abstract}
We show that the peak which can be observed in fidelity susceptibility around the Berezinskii-Kosterlitz-Thouless transition is shifted from the quantum critical point (QCP) at $J_c$ to $J^*$ in the gapped phase by a value $|J^* - J_c| = B^2/36$, where $B^2$ is a transition width controlling the asymptotic form of the correlation length $\xi \sim \exp(-B/\sqrt{|J-J_c|})$ in that phase. This is in contrast to the conventional continuous QCP where the maximum is an indicator of the position of the critical point. The shape of the peak is universal, emphasizing the close connection between fidelity susceptibility and the correlation length. We support those arguments with numerical matrix product state simulations of the one-dimensional Bose-Hubbard model in the thermodynamic limit, where the broad peak is located at $J^*=0.212$ that is significantly different from $J_c=0.3048(3)$. In the spin-$3/2$ XXZ model the shift from $J_c=1$ to $J^*=1.0021$ is small but the narrow universal peak is much more pronounced over the non-universal background.
\end{abstract}
\maketitle


{\it Introduction.---} 
A quantum phase transition occurs when a small variation of a parameter in a Hamiltonian leads to the dramatic change of the ground-state properties of the quantum system \cite{Sachdev}. This basic idea was behind suggesting fidelity -- the overlap between the ground states of the system for slightly shifted values of the external parameter $J$ --  as a universal probe of quantum criticality \cite{Zanardi2006}. 
Dramatic change of the system's properties across the critical point results in a drop of fidelity enabling both the location of the critical point and the determination of the universal critical exponent $\nu$ characterizing the divergence of the correlation length  \cite{Zanardi2006,zanardi_geometric,You2007,Gu2008,Chen2008,FS_Kitaev1,Zhou,Schwandt2009,ABQ2010,PolkovnikovArXiv2010,deGrandi2010a,*deGrandi2010b,Rams2011a,SenPRB2012,Zhou1,Adamski2013,Polk1}.
The fidelity has applications in as wide a context as that of the quantum phase transitions themselves.
It affects critical dynamics of decoherence \cite{FidDecoh}, has links with Fisher information and metrology \cite{Metrol,Rams_superHeisenberg_2018}, matters for shortcuts to adiabaticity \cite{UedaShort}, and is instrumental to define the geometry of quantum states \cite{PolkGeo}.

The natural approach to search for the critical point would be to fix the shift $\delta$ of the parameter $J$ and then scan various values of $J$. In the extreme limit of $\delta \to 0$ this is equivalent to looking at the fidelity susceptibility. 
This approach is well established by now \cite{GuReview,Troyer2015}. The position of a generic continuous critical point is indicated by the peak of the fidelity susceptibility defined as the second derivative of fidelity with respect to the small shift of the external parameter. An outstanding problem remains, however, in the case of the Berezinskii-Kosterlitz-Thouless (BKT) quantum phase transition and universal behavior of fidelity in its vicinity. We address this problem in this article. We show that in the BKT transition the value of the external parameter for which the system is changing most rapidly, manifested by the universal peak in fidelity susceptibility, is significantly shifted from the critical point toward the gapped phase. The shift is proportional to the width of the transition with a universal proportionality factor of $1/36$. 

There is a substantial body of literature on fidelity in the BKT transition \cite{Vezzani_PRL_2007,Yang_PRB_2007,Manmana1108,Rigol_BH_2013,Lacki_PRA_2014,Vekua2015}. 
Most of the results were obtained for finite system sizes.
The shifted peak was often seen, disbelieved, and attempts were made at its explanation.
For instance, 
Ref. [\onlinecite{Vekua2015}] argued that the peak is approaching the critical point logarithmically in the system size.
In this paper, in order to avoid any finite size effects, we consider the fidelity directly in the thermodynamic limit where the relevant quantity is a fidelity per lattice site:
\begin{equation}
\label{eq:def_f}
f(J_1,J_2) = - \lim_{N\to \infty }\frac {\log  \left |\langle J_1 | J_2 \rangle \right |}{N},
\end{equation}
with $|J_i \rangle$ being the ground state of Hamiltonian $H(J_i)$. The fidelity susceptibility follows as
\begin{eqnarray}
\label{eq:def_chif}
\chi_F(J) &=&  
\lim_{\delta \to 0} \frac{2 f(J-\frac{\delta}{2},J+\frac{\delta}{2})}{\delta^2} \\
  &=& \left.\frac{\partial^2 f(J-\frac{\delta}{2},J+\frac{\delta}{2})}{\partial \delta^2}\right|_{\delta=0}\nonumber,
\end{eqnarray}
which corresponds to expansion of the fidelity per site around its minimum at $\delta=0$, where $f(J,J)=0$, to the second order in $\delta$.

In the rest of the paper we discuss universal features of these quantities in the vicinity of the BTK critical point. We begin by briefly reviewing the universal scaling of fidelity in the vicinity of the conventional continuous critical point. We use these conventional results to show that for the BKT transition the maximum of fidelity susceptibility should appear deep in the gapped phase rather than at the critical point. Then we introduce a scaling ansatz for fidelity susceptibility, valid in the gapped phase around the BKT critical point, and use it to quantify the position of the maximum. We corroborate these predictions with infinite Density Matrix Renormalization Group (iDMRG) calculations for the Bose-Hubbard model and the spin-$3/2$ XXZ model, both in the thermodynamic limit.


{\it Generic scaling of fidelity susceptibility.---} 
Let us consider a Hamiltonian $H(J)$ in $d$ spatial dimensions with a continuous critical point at $J_c$. The external field $J$ is coupled to a relevant perturbation with a well defined scaling dimension.  
Near $J_c$ the correlation length diverges as $\xi \sim |J-J_c|^{-\nu}$, 
defining the critical exponent $\nu$.
Now, the universal contribution to fidelity susceptibility is expected to scale \cite{zanardi_geometric,Schwandt2009,ABQ2010,PolkovnikovArXiv2010,deGrandi2010a,*deGrandi2010b} as
\begin{equation}
\label{eq:fid_sus_away}
\tilde \chi_F (J) \propto |J-J_c|^{d\nu-2}.
\end{equation}
For instance, in the often considered exactly solvable \cite{Chen2008, Invernizzi08, Damski2013, Damski2014} one-dimensional Ising chain $\nu=1$ resulting in $\chi_F(J)\sim |J-J_c|^{-1}$. 
It is diverging for $J \to J_c$ and dominates the behavior of fidelity susceptibility when $d \nu <2$. Otherwise, non-universal system-specific corrections $\sim O(1)$ are dominant
and fidelity susceptibility cannot be used as a useful probe of the critical point. 

The above scaling predictions need to be carefully reconsidered in the vicinity of the BKT transition. In this case, when the system is tuned toward the critical point, the correlation length in the gapped phase is diverging faster than any polynomial:
\begin{equation}
\xi(J) \simeq \xi_0 \exp \left( B/ \sqrt{|J-J_c|} \right).
\label{eq:xi_KT}
\end{equation}
Consequently, Eq.~\eqref{eq:fid_sus_away} cannot be directly used. Above, we can interpret $B^2$ as a non-universal width of the transition. 

A heuristic way to proceed is to locally approximate the exponential divergence in Eq.~\eqref{eq:xi_KT} with a power law. This introduces an effective exponent: $\nu(J) = \partial \log \xi(J) / \partial \log |J - J_c| = \frac{B}{2}|J-J_c|^{-1/2}$. Now $\nu(J)$ diverges as $J\to J_c$, suggesting that the universal contribution in Eq.~\eqref{eq:fid_sus_away} is subleading in this limit. However, for large enough $|J-J_c|$ there is a regime where $\nu(J) < 2/d$ and the universal contribution can dominate. This opens a possibility that a peak of fidelity appears in that regime for some value of $J^*$ different than $J_c$. Its position depends on the width $B^2$, with a smaller width leading to a more pronounced peak at $J^*$ closer to $J_c$.


{\it  Scaling ansatz.---}  In order to quantify this general intuition we follow \cite{Rams2011a,Rams2011b} and postulate the scaling hypothesis for the universal part (we introduce a tilde to indicate this) of 
log-fidelity per site $\tilde f(J_1,J_2) = b^{-d} g\left (\xi(J_1) b^{-1},\xi(J_2) b^{-1}\right)$.

A similar approach was also, for instance, very recently used to characterize the behavior of the Loschmidt echo in the vicinity of the (conventional) continuous critical point~\cite{Hwang_Loschmidt_2019}.
Using the freedom to choose the scaling factor we can rewrite it as
 \begin{equation}
\label{eq:KT_scaling}
\tilde f(J_1,J_2) =  \xi(J_1)^{-d} g\left (1,\xi(J_2)/\xi(J_1)\right).
\end{equation}
The second factor has a minimum equal $0$ for $J_1=J_2$, i.e., $g(1,1) =0$. There are two interesting limits to consider here, both revealing universal behavior.

First, when $J_2=J_c$ is tuned exactly to the critical point and $J_1 = J_c + \delta$ is away from the critical point in the gapped phase. This allows one to conclude that
\begin{equation}
\label{eq:fc}
\tilde f(J_1,J_c) \sim \xi(J_1)^{-d}.
\end{equation}
For the conventional critical point with $\xi(J_c+\delta) \sim |\delta|^{-\nu}$, Eq.~\eqref{eq:fc} leads \cite{Rams2011a} to the scaling $f(J_c+\delta,J_c) \sim |\delta|^{d\nu}$. Here we extend this prediction to the BKT transition. By employing Eq.~\eqref{eq:xi_KT} and setting $d=1$, we obtain 
\begin{equation}
 \tilde f(J_1,J_c) \sim \exp \left(-B/\sqrt{|J_1-J_c|} \right).
 \label{eg:fid_critical}
\end{equation}
Such contribution is vanishing exponentially for $J_1 \to J_c$. However, depending on the smallness of $B$, for some intermediate range of $\delta$ it might still be visible above the non-universal background $\sim O(|J_1-J_c|^2) = O(\delta^2)$. In the following we support this with numerical results. 

Second, we can set $J_{1,2} = J\pm \frac{\delta}{2}$ and Taylor expand $\tilde f(J_1,J_2) $ to the second order in $\delta$. This gives the fidelity susceptibility as a second derivative of Eq.~\eqref{eq:KT_scaling} calculated at $\delta=0$ [see Eq.~\eqref{eq:def_chif}]. Note that $\tilde f(J_1, J_2)$ has a minimum for $J_1 = J_2$ when fidelity is calculated with respect to the same state, i.e., for $\delta =0$, and consequently $g(1,1) = g'(1,x)|_{x=1} = 0$.
For the conventional critical point, $\xi \sim |J-J_c|^{-\nu}$, this gives an alternative derivation of scaling of fidelity susceptibility in Eq.~\eqref{eq:fid_sus_away} (see Ref. [\onlinecite{Rams2011a}]).

Here, we extend this analysis to the BKT transition. Employing Eq.~\eqref{eq:xi_KT} and Taylor expanding Eq.~\eqref{eq:KT_scaling} to the second order we obtain the scaling of the universal part of fidelity susceptibility as
\begin{equation}
\tilde \chi_F(J) \simeq A B^2 \frac{ \exp( -B/\sqrt{|J-J_c|}) }{|J-J_c|^3},
\label{eq:fid_KT_away}
\end{equation}
with a prefactor $A = \xi_0^{-1} g''(1,x)|_{x=1}/4$. 
It provides a good approximation for the log-fidelity per site $\tilde f(J-\frac{\delta}{2},J+\frac{\delta}{2})\simeq \frac{1}{2}\delta^2 \tilde \chi_F(J)$ for $\delta \ll |J-J_c|^{3/2}$, i.e. in the limit of two states being close as compared with their distance from the critical point. Otherwise higher orders in the expansion in $\delta$ should become relevant, [see, e.g., Eq.~\eqref{eg:fid_critical}].

Equation~\eqref{eq:fid_KT_away} is the central prediction of this paper.
The maximum of the above universal contribution is reached for $J^*$ such that
\begin{equation}
|J^*-J_c| = B^2/36.
 \label{eq:Jmax}
\end{equation}
It is significantly shifted  from the critical point, by a value proportional to the system-specific transition width $B^2$. The proportionality factor, $1/36$, is universal, however. The magnitude of the peak scales as $B^{-4}$ or, more precisely, $115.6 A B^{-4}$  with other parameters fixed. The peak is more pronounced for a more narrow transition. In the following we consider two models with widely different widths $B^2$.


\begin{figure} [t!]
\begin{center}
  \includegraphics[width= \columnwidth]{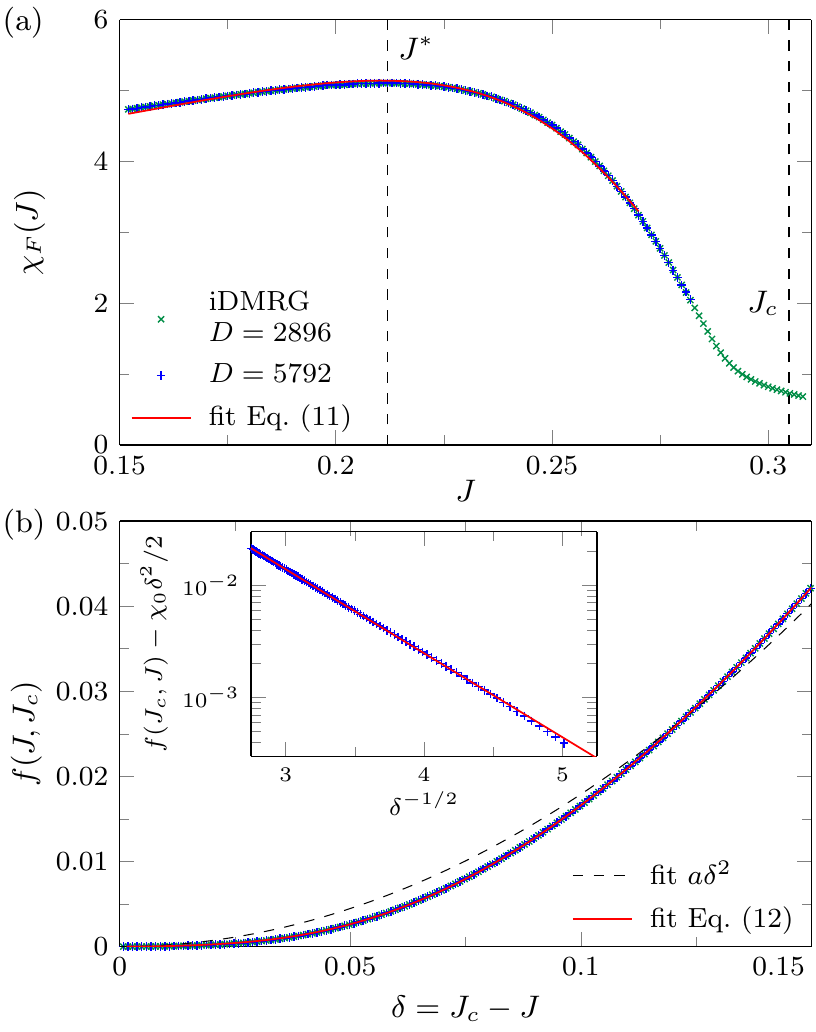}
\end{center}
  \caption{ 
    Fidelity in the Bose-Hubbard model at unit filling $\langle n_m \rangle = 1$.
    In (a)
    we show the fidelity susceptibility (points), calculated using $\delta = 0.004$. The broad maximum is found at $J^* \simeq 0.212$, whereas the position of the critical point~\cite{Rams_PRX_2018} is found at $J_c \simeq 0.3048$. We fit (red line) the dependence in Eq.~\eqref{eq:fid_sus_full} and obtain: $\overline \chi_0 = 2.00(5)$, $A=0.312(4)$, $B = 1.84(5)$. The error bars correspond to the fitting error. 
    In (b) 
    we show the fidelity per site calculated as an overlap with the critical point (points). We fit (red line) Eq.~\eqref{eg:fid_critical_fit}. The fit gives $\chi_0=1.23(2)$, $B=1.72(1)$. In the fit, the position of the critical point was fixed at $J_c = 0.3048$.
    Finally, the inset shows the difference between the fidelity and the non-universal background emphasizing the universal part. Red line is plotted using parameters obtained above by fitting the total fidelity in the linear scale, showing full consistency with Eq.~\eqref{eg:fid_critical_fit}.
    We show the results for two values of the uMPS bond dimensions ($D=2896$ and, focusing on the peak, $D=5792$) to indicate very good convergence of the obtained numerical data. In iDMRG simulations local Hilbert space was truncated at $n_m=6$ particles.
  }
  \label{fig:BH_xi}
\end{figure}


{\it Bose-Hubbard model.---}
First we consider
\begin{equation}
H(J) = - J \sum_m  \left (   b_{m+1}^\dagger   b_m +   b_{m}^\dagger   b_{m+1} \right) + \frac{U}{2} \sum_m     n_{m} (  n_{m} - 1) ,
\end{equation}
where $b_m$ is a bosonic annihilation operator on site $m$ and $n_m = b_m^\dagger b_m$ is a particle number operator. We set the energy scale by fixing $U=1$. We consider a unit filling per lattice site, $\langle  n_{m} \rangle = 1$, in which case the model exhibits a quantum phase transition in the  BKT universality class \cite{Fisher_BH_1989,Krutitsky_BH_review_2016} between a gapped Mott insulator phase for $J<J_c$ and a gapless superfluid phase for $J>J_c$.

We employ uniform matrix product states (uMPS) simulations of the model \cite{verstraete2008matrix,schollwock2011,orus_review_2014,McCulloch_idmrg_2008,Singh_sym_2010,Singh_U(1)_2011} taking advantage of the fact that the fidelity per lattice site can be directly calculated from the largest eigenvalue of the mixed transfer matrix naturally occurring in a scalar product between two uMPS. We employ a variant of the iDMRG algorithm \cite{McCulloch_idmrg_2008} with a two-site unit cell incorporating $U(1)$ symmetry \cite{Singh_sym_2010,Singh_U(1)_2011}. For the Bose-Hubbard model it corresponds to conservation of the total particle number. Employing symmetries not only greatly speeds up the simulations allowing one to reach significantly larger uMPS bond dimensions, but also lets us fix the desired particle density without the need to resort to chemical potential. All simulated states were converged up to maximal change of Schmidt values in the last iteration below $10^{-10}$.

We find the position of the critical point, as well as the reference value of the parameter $B$ from the divergence of the correlation length in Eq.~\eqref{eq:xi_KT}.
Precise extraction of the correlation length from uMPS requires proper extrapolation, as was recently shown in Ref.~[\onlinecite{Rams_PRX_2018}] by some of us. Correlation length is a non-local quantity which converges very slowly with uMPS bond dimension $D$. Proper extrapolation however, allows one to effectively take the limit $D\to \infty$. In that article, the correlation function was fitted with the scaling form 
$ \log \xi = \log \xi_0 + B/\sqrt{J_c-J}  + a_2\sqrt{J_c-J} $, which also includes a sub-leading correction. For the Bose-Hubbard model considered here, this leads to $J_c = 0.3048(3)$, $B=1.61(4)$, $a_2= -3.52(24)$ and $\xi_0 = 0.262(39)$. These values of $J_c$ and $B$ are in very good agreement with an independent fit of the scaling of the energy gap in a finite system in Ref.~[\onlinecite{Rigol_BH_2013}]. A number of numerical estimates of $J_c$ obtained within various previous studies (including among others the position of the maximum of the fidelity susceptibility obtained in Ref.~[\onlinecite{Vezzani_PRL_2007}] for a finite system of few sites) is collected in Table I in the review [\onlinecite{Krutitsky_BH_review_2016}].

Numerical simulations of fidelity susceptibility in the thermodynamic limit are shown in Fig.~\ref{fig:BH_xi}(a) with the wide maximum located at $J^*=0.212$.
In order to fit numerical data we supplement the universal prediction (\ref{eq:fid_KT_away}) with a non-universal sub-leading constant $\overline \chi_0$:  
\begin{eqnarray} \label{eq:fid_sus_full}
\chi_F(J) \simeq  \overline \chi_0 +  AB^2 \frac{ e^{-B |J-J_c|^{-1/2}} }{  |J-J_c|^{3}  }  .
\end{eqnarray}
Above, $\overline \chi_0$ can be understood as averaging the unknown non-universal contribution over the considered window of $J$'s. Note that the latter can in principle also depend on $J$. To minimize such effect, without introducing a more complicated fitting model, we focus the fit on the vicinity of the peak. The value of $\xi_0 =2.00(5)$ (see Fig.~\ref{fig:BH_xi}) and $\xi_0 =1.23(2)$ calculated below at the critical point differ slightly which we expect is resulting from slightly varying non-universal contributions. Those changes are however, a few times smaller than the size of the peak at $J^*$ and for smooth changes should not much affect the position of the peak.

By the same token, the universal part of the fidelity per site with respect to the critical point in Eq. (\ref{eg:fid_critical}) is supplemented with a non-universal background:
\begin{equation}
 f(J,J_c) \simeq \frac12 \chi_0 \left(J-J_c\right)^2 + A_0 \exp \left(-B/\sqrt{|J-J_c|} \right),
 \label{eg:fid_critical_fit}
\end{equation}
with $A_0 = g(1,\infty)$.
Numerical simulations are shown in Fig.~\ref{fig:BH_xi}(b). As expected, the non-universal background dominates over the universal part close enough to the critical point, but adding the universal part in Eq.~\eqref{eg:fid_critical_fit} is necessary to describe the observed data [compare with the dashed line in Fig.~\ref{fig:BH_xi}(b)]. The inset shows that in the intermediate range of $J_c-J$ the universal part can be discerned from the background and is captured by the fit with remarkable accuracy. This demonstrates self-consistency of our scaling theory.


\begin{figure} [t!]
\begin{center}
  \includegraphics[width= 1\columnwidth]{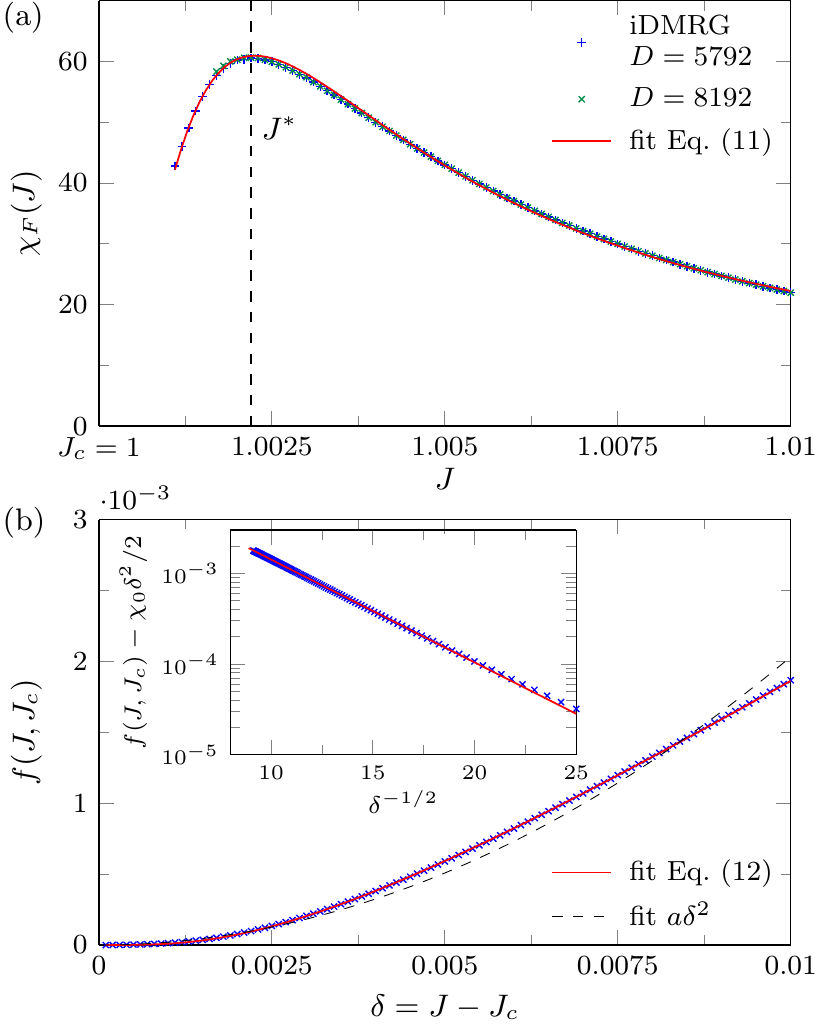}
\end{center}
  \caption{ 
  Fidelity in XXZ spin$-3/2$ model. 
  In (a) 
  we show the fidelity susceptibility (blue), calculated using $\delta = 0.001$. The narrow maximum is found at $J^* \simeq 1.0021$, wheres the position of the critical point is found at $J_c =1$. We fit (red line) the dependence in Eq.~\eqref{eq:fid_sus_full} and obtain $\overline \chi_0 = 8.2(3)$, $A=0.0030(1)$, $B = 0.285(1)$. The error bars corresponds to the $95\%$ confidence bounds from the nonlinear fit.
  In (b) 
  we show the fidelity per site calculated as an overlap with the critical point (blue). We fit (red line) Eq.~\eqref{eg:fid_critical_fit}. The fit gives $\chi_0=8.9(3)$, $B=0.261(1)$. The inset shows the difference between the fidelity and the non-universal background emphasizing the universal part, where the red line is plotted using the data fitted above for the total fidelity.
  }
   \label{fig:XXZ_fs}
\end{figure}


{\it XXZ spin-$\frac32$ model.---}
Next we proceed with
\begin{equation}
\label{eq:XXZ32}
H = \sum_m  \left(   S^x_m  S^x_{m+1} +  S^y_m  S^y_{m+1}  + J  S^z_m S^z_{m+1} \right),
\end{equation}
where $S^{x,y,z}_m$ are standard spin-$\frac32$ operators acting on site $m$. The model has the BKT critical point \cite{XXZ32_1986_Schulz,Affleck_1987,Affleck_1989,XXZ32_Hallberg_1996} at $J_c = 1$ separating the gapped phase for $J > 1$ from the gapless region for $-1<J<1$. Reference [\onlinecite{Rams_PRX_2018}] reports fitting the scaling form of the correlation length as
$ \log \xi = \log \xi_0 + B/\sqrt{J_c-J}  + a_2\sqrt{J_c-J} $, with $B= 0.304(12)$, $a_2=-5.4(12)$ and $\xi_0=4.26(85)$. The fitted $J_c =0.99993(4)$ obtained there is in excellent agreement with the exact value of $J_c=1$.
The relative transition width $B^2/J_c$ is almost eight times smaller than in the BH model making the fidelity susceptibility peak more pronounced at $J^*$ that is closer to $J_c$. 

We use Eqs. (\ref{eq:fid_sus_full},\ref{eg:fid_critical_fit}) to fit numerical results for the fidelity susceptibility and the fidelity per site with respect to the critical point. The results of our numerical simulations are shown in Fig.~\ref{fig:XXZ_fs}. The fitting parameters are consistent with those obtained from the correlation length. Again, the scaling theory is able to discern the universal part of the fidelity with respect to the critical point from its non-universal background.

{\it Conclusion.---}
The shifted fidelity peak is not an artifact but hard reality. When measured by a distance in the Hilbert space, the ground state undergoes the fastest changes not at the BKT critical point but away from it in the gapped phase. The shape of the peak is universal, just as universal is its shift equal to $1/36$ of the width of the transition.

\acknowledgments

This research was funded by the National Science Centre (NCN), Poland under projects 2016/23/D/ST3/00384 (MMR) and 2016/23/B/ST3/00830 (JD), 
and DOE under the Los Alamos National Laboratory LDRD Program (LC, WHZ). One of us (WHZ) was funded in part by the National Science Foundation under Grant No. NSF PHY-1748958 while at KITP, UCSB. LC
was also supported by the DOE through the J. Robert Oppenheimer fellowship.


\input{final.bbl}

\end{document}

%% file: final.bbl
%